\documentclass[twocolumn,preprintnumbers,amsmath,amssymb]{revtex4}
\usepackage[dvipdfmx]{graphicx}
\usepackage{dcolumn}
\usepackage{bm}
\usepackage{multirow}

\begin{document}

\title{Measurement of the variation of electron-to-proton mass ratio using
ultracold molecules produced from laser-cooled atoms}

\author{J. Kobayashi}
\affiliation{Department of Physics, Graduate School of Science, Kyoto University, Kyoto 606-8502, Japan}\affiliation{PRESTO, Japan Science and Technology Agency, Kyoto 606-8502, Japan}
\email{j-kobayashi@scphys.kyoto-u.ac.jp}
\author{A. Ogino}
\affiliation{Department of Applied Physics, University of Tokyo, Bunkyo-ku, Tokyo 113-8656, Japan}
\author{S. Inouye}
\affiliation{Graduate School of Science, Osaka City University, Osaka 558-8585, Japan}
\affiliation{Nambu Yoichiro Institute of Theoretical and Experimental Physics (NITEP),
Osaka City University, Osaka 558-8585, Japan}

\begin{abstract}
Experimental techniques to manipulate cold molecules have seen great development in recent years. The precision measurements of cold molecules are expected to give insights into fundamental physics.
We use a rovibrationally pure sample of ultracold KRb molecules to improve the measurement on the stability of electron-to-proton mass ratio ($\mu = \frac{m_{\rm e}}{M_{\rm p}}$).
The measurement is based upon a large sensitivity coefficient of the molecular spectroscopy,
which utilizes a transition between nearly a degenerate pair of vibrational levels each associated with a different electronic potential.
Observed limit on temporal variation of $\mu$ is $\frac{1}{\mu}\frac{d\mu}{dt} =  (0.30\pm1.0) \times 10^{-14}$\,year$^{-1}$,  which is better by a factor of five compared with the most stringent laboratory molecular limits to date.
Further improvements should be straightforward, because our measurement was only limited by statistical errors.

\end{abstract}
\date{\today}
\maketitle

Cold molecules are becoming a popular tool for precision measurements.
The search for the  electron electric dipole moment (EDM) is a good example\cite{ACME18, JILAEDM17, hind11}.
Historically, hot atoms in a vapor cell or in an atomic beam were gradually replaced with
cold molecular sources and trapped molecular ions.
Cold molecules have a number of advantages for increasing the precision, like
longer interaction time, smaller motional decoherence,
 and it's easier to make use of its rich internal degrees of freedom\cite{carr09}.
In the case of  EDM experiments, molecules (and molecular ions) were placed in $\Omega$ doublet levels, where large internal electric field and systematic error rejection are available\cite{demi13}.
Similarly, one can expect to improve the measurement of the variation of fundamental constants by selecting molecules with a level structure that enhances the sensitivity.

Thus far, various physical systems and experimental techniques have been used to measure the stability of the  electron-to-proton mass ratio {\bf $\mu = m_{\rm e}/M_{\rm p}$}: each of them provide a crucial independent check.\cite{uzan03,demi17}. 
Molecular spectroscopy is often called model-free, because it directly reflects the stability of the inertial mass of the nucleus.
Shelkovnikov {\it et al.} studied the absorption lines of SF$_6$ and obtained a limit on the current variation of $\mu$ as $\frac{1}{\mu}\frac{d\mu}{dt} = (-3.8\pm5.6) \times 10^{-14}\,\rm{year}^{-1}$\cite{shel08}. 

In the mean time, impressive results of
$\frac{1}{\mu}\frac{d\mu}{dt} = 0.2(1.1) \times 10^{-16}\,\rm{year}^{-1}$ \cite{godu14} and
$\frac{1}{\mu}\frac{d\mu}{dt} = -0.5(1.6) \times 10^{-16}\,\rm{year}^{-1}$ \cite{hunt14}
were obtained by using atomic clocks.
These measurements are essentially measuring the variation of the electron-to-proton magnetic moment ratio.

Here we report a measurement on the variation of {\bf $\mu$} using ultracold KRb molecules. Diatomic alkali molecules like KRb have two low-lying, overlapping potentials:
the deep $X^1\Sigma^+$ ground state and the shallower $a^3\Sigma^+$ state.
It has been pointed out that using a transition between
a nearly degenerate pair of vibrational levels each associated with {\bf $X^1\Sigma^+$} and {\bf$a^3\Sigma^+$}
 potentials,
one can realize an enhanced sensitivity to the variation of $\mu$ \cite{demi08}.
We realize such an enhancement (on the order of $\sim 10^4$) by using ultracold KRb molecules produced
by photoassociation in a Magneto-Optical Trap followed by Stimulated Raman Adiabatic Passage \cite{aika10}.
We obtain a better result than previous measurements using molecules \cite{shel08}.
That is solely limited by counting statistics and demonstrates the power of the method.

\begin{figure}
\includegraphics[width=9cm]{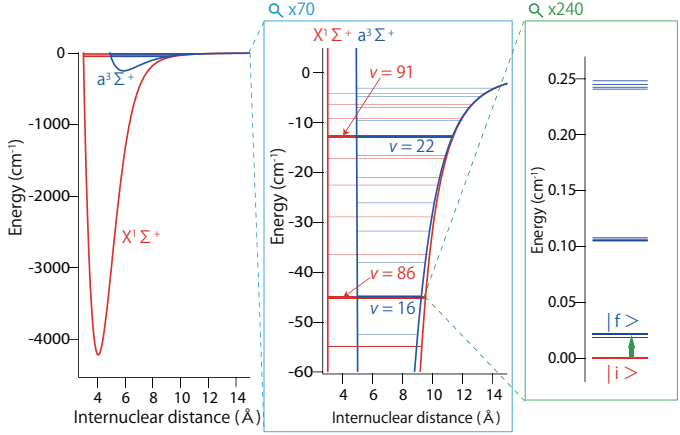}
\caption{\label{fig:XandA}
Closely lying states of a $^{41}$K$^{87}$Rb molecule\cite{pash07}.
In the shallow range of the potentials, several pairs of closely lying vibrational states in $X^1\mathrm{\Sigma^+}$ and $a^3\mathrm{\Sigma^{+}}$ potentials are found.
Among such pairs, we selected the combination of $v = 86$ state in $X^1\mathrm{\Sigma^+}$ and $v = 16$ state in $a^3\mathrm{\Sigma^{+}}$. 
The large difference of potential depth between $X^1\mathrm{\Sigma^+}$ and $a^3\mathrm{\Sigma^{+}}$ leads to the large sensitivity $W\equiv\partial\nu/\partial(\mathrm{ln} \,\mu)\approx -9.5$ THz, whereas the transition frequency itself ($\equiv\nu$) is in the microwave range ($\approx 635$ MHz).
This yields a large sensitivity coefficient $|K_\mu|\equiv|{W}/{\nu}| \approx 15000$.
}
\end{figure}

\section*{Results}
\subsection*{Sensitivity to the variation of $\mu$}

To quantify the sensitivity of the transition to the variation of fundamental constants, we define the sensitivity $W_{\mu}$ and the sensitivity coefficient $K_{\mu}$ as
\begin{equation}
{\Delta \nu}=W_{\mu} \frac{\Delta \mu}{\mu}= \nu K_{\mu} \frac{\Delta \mu}{\mu} .
\end{equation}
$W_{\mu}$  and $K_{\mu}$ determine the absolute and relative accuracy  needed for frequency standards used in the experiment.  
In general, $|K_\mu|$s for vibrational and rotational transitions of diatomic molecules are 1/2 and 1, respectively.
Large sensitivity coefficients can be realized using transitions between nearly degenerate states with different symmetries. Such transitions were successfully used in previously mentioned astronomical observations, where mixed torsion-rotation transitions with $K_\mu = -32.8$ were investigated\cite{bagd13}. A similar idea was used to improve the limit on the variation of the fine-structure constant\cite{leef13}.

\subsection*{Target transition}
In this study, we used such transitions in ultracold KRb molecules. As pointed out in Ref. \cite{demi08}, closely lying states in $X^1\mathrm{\Sigma^+}$ and  $a^3\mathrm{\Sigma^{+}}$ potentials give rise to a large $K_\mu$.
Figure \ref{fig:XandA} shows the actual energy curves calculated for $^{41}\mathrm{K}^{87}\mathrm{Rb}$ \cite{pash07}. 
From these potential energy curves we can calculate the sensitivity for each vibrational state with the 
accuracy only being limited by the uncertainty of the potential energy curve(Method).
Although the vibrational states with the binding energies about 1000 cm$^{-1}$ in $X^1\mathrm{\Sigma^+}$
have the highest sensitivities of $W_{\mu}$ = 35 THz, their energies are far from that of the vibrational states in $a^3\mathrm{\Sigma^{+}}$, whose largest binding energy is about 240 cm$^{-1}$. 

We found that one of the best combinations is that of the $v = 86$ state in the $X^1\mathrm{\Sigma^+}$ potential and the $v=16$ state in the $a^3\mathrm{\Sigma^{+}}$ potential because the transition frequency is in the microwave range, the sensitivity coefficient is large ($\sim 10^4$), and the strength of the transition is not too small owing to the finite singlet-triplet mixing ($\sim 2\%$).
Though its sensitivity is not the highest, it is as high as about one fourth of that.
Moreover, it is another advantage that the experiments with microwaves are easier than the experiments with frequencies in the optical domain. 

\begin{figure}
\includegraphics[width=9cm]{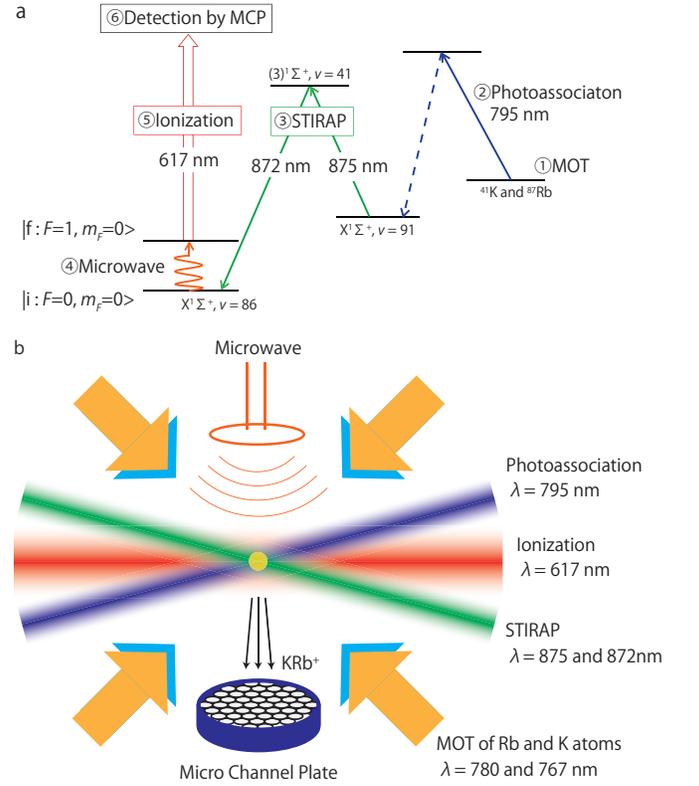}
\caption{\label{fig:schematic}
Production and detection of ultracold molecules.
First, ultracold atoms in a dual-species magneto-optical trap (MOT) were photoassociated into
a weakly bound state. The molecules were transferred to the target internal state by stimulated Raman adiabatic passage (STIRAP) and irradiated with a microwave pulse.
State selective detection of the molecules was achieved by ionizing the molecules with a pulsed laser and detecting them with a microchannel plate (MCP). }
\end{figure}

Understanding the details of the closely lying hyperfine manifolds is mandatory for estimating the strength of each microwave transition.
In the following discussion, we express the hyperfine states as $|S, F_1, F, m_F\rangle$ in short, where the detail of labeling is in the Method section. 
Note that $S=0$ ($S=1$) signifies the $v = 86$ state in $X^1\mathrm{\Sigma^+}$  ($v = 16$ state in $a^3\mathrm{\Sigma^{+}}$).

We obtained the hyperfine spectra by sweeping the frequency of the STIRAP laser (Fig. 2).
The results are summarized in the Supplementary note 1 and Supplementary table 1.
We identified the most suitable transition to test the stability of $\mu$ as $|\mathrm{i}\rangle$-$|\mathrm{f}\rangle$ (Fig.1) where
\begin{eqnarray*}
|\mathrm{i}\rangle & \equiv& |S = 0, F_1 = 3/2, F =0, m_F = 0\rangle,\\
|\mathrm{f}\rangle & \equiv& |S = 1, F_1 = 1/2, F = 1, m_F = 0\rangle.
\end{eqnarray*}
The $|\mathrm{i}\rangle$-$|\mathrm{f}\rangle$ transition fulfills the required conditions: the sensitivity to the variation of $\mu$ is large, the magnetic sublevels are resolved by a small magnetic field, and there is no first-order Zeeman shift. The transition frequency is $\nu_{\rm i-f}= 634.96 \,{\rm MHz}$, and the sensitivity of the transition to the variation of $\mu$ is $W_{\rm i-f}=\partial \nu_{\rm i-f}/\partial(\mathrm{ln}\mu)= -9.45(4){\rm THz}$(Methods).
The resulting sensitivity coefficient is
\begin{equation}
K_{\mu, {\rm i-f}} =\frac{W_{\rm i-f}}{\nu_{\rm i-f}} =  -14890(60).
\end{equation}

\subsection*{Spectroscopy}
\begin{figure}
\includegraphics[width=9cm]{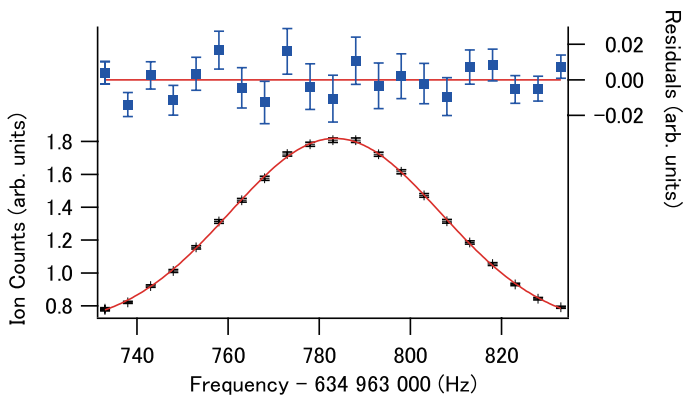}
\caption{\label{fig:gaussian} 
Typical spectrum of the $|\mathrm{i}\rangle$ - $|\mathrm{f}\rangle$ transition.
Average of the ion signals and its standard deviations of mean (shown in black) are plotted against the microwave frequency.
The total number of original data points used to make this plot is approximately 95000, which corresponds to 6 hours of data collection.
To find the central frequency, the spectrum is fitted to a Gaussian function (red curve).
Subtracting the Gaussian function from the ion counts gives the clear view of the residuals.
The full-width-of-half-maximum (FWHM) of the spectrum is about 50 Hz, which is limited by the microwave pulse duration of 16 ms.
}
\end{figure}

Figure \ref{fig:gaussian} shows a typical spectrum of the $|\mathrm{i}\rangle$ - $|\mathrm{f}\rangle$ transition obtained by six hours of data collection. A Gaussian fit to the data provided the full-width-of-half-maximum(FWHM) of approximately 50 Hz, which is consistent with the ideal spectrum obtained by the microwave $\pi$-pulse of 16 ms duration.
Longer pulse durations degraded the signal because
ballistically expanding clouds of molecules with mean velocity $\sim 130$ mm\,s$^{-1}$ started to leave the area of the ionization beam whose radius was approximately 2 mm.
To access only the $m_F=0$ state in the $|S=1, F_1 = 1/2, F = 1\rangle$ manifold a magnetic field of approximately 1.5 ${\rm{\mu}}$T was applied.
The measurements were repeated every 120 ms.
The statistical uncertainty of the central frequency was approximately 93 mHz.

\subsection*{Statistical and systematic errors}
We paid special attention to the magnetic field during the measurement as it drifts because of the magnetization of the metal chamber and fluctuates due to other environmental factors.
The $|\mathrm{i}\rangle$ - $|\mathrm{f}\rangle$ transition is immune to the first-order Zeeman shift; however, it is affected by the second-order Zeeman shift ($\sim0.82 \,{\rm Hz}\,{\mu}$T$^{-2}$).
To evaluate this shift, we simultaneously measured the transition frequency to the hyperfine state
$|\mathrm{f_1}\rangle\equiv|S= 1, F_1 = 1/2, F = 1, m_F = 1 \rangle$ whose energy is sensitive to the first-order Zeeman shift ($\sim3.3 \,{\rm kHz}\,{\mu}$T$^{-1}$).
The observations for $|\mathrm{i}\rangle$ - $|\mathrm{f}\rangle$ and $|\mathrm{i}\rangle$ - $|\mathrm{f_1}\rangle$ alternated every 2.4 s, and the frequency of the $|\mathrm{i}\rangle$ - $|\mathrm{f_1}\rangle$ transition was used to subtract the second-order Zeeman shift from the measured $|\mathrm{i}\rangle$ - $|\mathrm{f}\rangle$ transition frequencies.
The statistical and systematic errors caused by the correction step were evaluated as 15 mHz and 1 mHz, respectively, where the systematic error is caused by the error of the Zeeman coefficients.

The blackbody radiation (BBR) shift can fluctuate owing to the temperature fluctuation of the vacuum chamber ($26.8\pm2 ^\circ$C). The coefficient was calculated using {\it ab initio} calculations for the transition dipole moments \cite{beuc06}.
The BBR shifts for $|\rm{i}\rangle$ and $|\rm{f}\rangle$ states are on the order of 1 Hz, but the effect on the transition frequency is far smaller since two shifts almost cancel out.
Because of the difficulty of the calculation, the uncertainty of the BBR shift on the transition frequency is larger than the central value of the estimation, where the BBR shifts induced by the coupling with other vibrational states in $X^1\Sigma^+$ and $a^3\Sigma^{+}$ are much smaller than that induced by the coupling with other electronic states.
From both of the temperature fluctuation and the calculated BBR coefficients with the uncertainty (0.08(20) Hz at 300 K), we estimated the fluctuation of the blackbody shift during the measurements as less than 10 mHz.

There are several systematic effects with fluctuations less than 1 mHz; hence, we can neglect them in the uncertainty budget table (Table \ref{tbl:error_budget}).
The list includes the Stark shift, the density shift, and the reference clock.
The Stark shift is caused by a static electric field in the vacuum chamber, which effectively shields  environmental electric field fluctuations.
The main source of the electric field in the detection region is the microchannel plate (MCP). We canceled this field during the measurements by applying a compensation field from electrodes placed inside the chamber, where we measured the electric field by observing the microwave transition between the rotational states of $|X^1\Sigma^+, v=0, N=0\rangle$ and $|N=1\rangle$, whose sensitivity to the electric field ($\sim12.4 \,{\rm Hz}\,\rm{cm}^2\,\rm{V}^{-2}$) is one order higher than that of the $|\mathrm{i}\rangle$ - $|\mathrm{f}\rangle$ transition ($\sim1.08 \,{\rm Hz}\,\rm{cm}^2\,\rm{V}^{-2}$).
However, a finite residual electric field of 0.19(6) V cm$^{-1}$ persists and it causes a dc-Stark shift of approximately 40 mHz, where the stark shift coefficients are estimated by using the {\it ab initio} calculation for the dipole moment \cite{stark}.
This issue, however, is not important because the fluctuation of the Stark shift, which comes from the fluctuation of the MCP voltage, should be less than 1 mHz.
The fluctuation of the frequency clock is also negligible because we use a commercial Rb-frequency standard (SRS FS725) locked to the Global Positioning System signal ($\delta f/f\sim10^{-12}$).
The density shift is also less than 1 mHz because we blast off the atomic clouds by shining the resonant lasers before the microwave application. 
Note that though the photoassociation recoil induces a finite mean velocity to the molecular cloud and that causes a frequency shift of about 10 mHz at most by the Doppler effect for microwave, the shift does not affect the measurement of the temporal variation of $\mu$, because all measurements were performed in the identical configuration and the shift is also identical for all measurements.

\begin{table}
\begin{center}
\begin{tabular}{lr}
Statistical		& Uncertainty  \\ \hline
$|\mathrm{i}\rangle$ - $|\mathrm{f}\rangle$ transition & 93 mHz \\
2nd-order Zeeman	& 15 mHz \\ \hline
Statistical total & 94 mHz \\ \hline \hline
 & \\
 &\\
 Systematic      		& Fluctuation \\ \hline
 2nd-order Zeeman   		&  1 mHz \\ 
 Blackbody radiation 		&  10 mHz \\
 Stark       				&  1 mHz \\
 Reference clock &  $<$ 1 mHz\\
 Density shift				& $<$ 1 mHz \\ \hline
 Systematic total & 10 mHz\\ \hline\hline
\end{tabular}
\caption{\label{tbl:error_budget} Error budget table for the measurement in Fig. \ref{fig:gaussian}.
The total uncertainty is dominated by the statistical error of the $|\mathrm{i}\rangle$ - $|\mathrm{f}\rangle$ transition measurements.
}
\end{center}
\end{table}

\begin{figure}
\includegraphics[width=9cm]{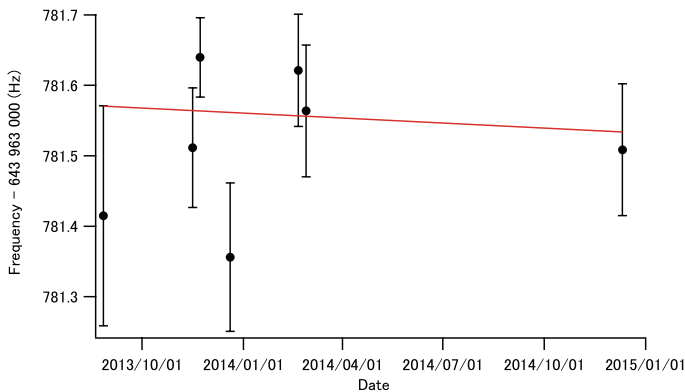}
\caption{\label{fig:temporal} Measurement of the temporal variation of $\mu$. The resonance frequencies of the $|\mathrm{i}\rangle$ - $|\mathrm{f}\rangle$ transition are plotted for compensated Zeeman shifts. The error bars denote statistical standard deviations of mean. The red line shows a linear fit to the data with statistical errors. We obtained the temporal variation of the transition frequency $f$: $\frac{1}{f}\frac{df}{dt} = (-0.44\pm1.47_{\mathrm{stat}}\pm0.24_{\mathrm{syst}})\times10^{-10}$\, \rm{year}$^{-1}$.}
\end{figure}

\subsection*{Stability of $\mu$}
We accumulated data intermittently for about 16 months and present the result in Figure \ref{fig:temporal}.
From the linear fit to the data, we obtain the temporal variation of $f$ as 
\begin{equation}
\frac{1}{f}\frac{df}{dt}= (-0.44\pm1.47_{\rm stat}\pm0.24_{\rm syst})\times10^{-10}\, \rm{year}^{-1}
\end{equation}
and the temporal variation of $\mu$ as
\begin{equation}
\frac{1}{\mu}\frac{d\mu}{dt}= (0.30\pm1.00_{\rm stat}\pm0.16_{\rm syst})\times10^{-14}\, \rm{year}^{-1}.
\end{equation}
This is the most accurate molecular test for the temporal variation of $\mu$ to date.
In a precise discussion, we measured the frequency ratio of $f/f_{\rm{Cs}}$, where the $f_{\rm{Cs}}$ is the clock transition frequency of Cs atom. However, since the precision of our measurement is about $10^{-10}$, which is much lower than that of the other stability tests using $f_{\rm{Cs}}$, we ignored the shift of $f_{\rm{Cs}}$ for simplicity.

\section*{Discussion}
We have derived a molecular limit on the temporal variation of $\mu$ by observing the microwave transition of photoassociated KRb molecules.
The result is noteworthy for the following reasons.
First, because of the large sensitivity coefficient of about $10^4$ the variation of $\mu$ was evaluated with four orders of magnitude higher accuracy than that of the frequency in our measurement.
This can be a great benefit to the improvement of the precision spectroscopy because we can greatly reduce the necessity of the precision of the reference clock.
Second, the sensitivities to the variation of $\mu$ of the molecular levels can be calculated from the experimentally determined molecular energy curves, thus they can be determined almost model independently.  
Improving the measurements using the molecular levels is important because it is independent of the measurement related to the atomic nuclear magnetic moments.
Third, since ultracold molecules synthesized from ultracold atoms are used, the application to the next generation of spectroscopic methods is expected. 
We started to construct a machine to incorporate the idea in Ref.\cite{koba14} to cool alkali-metal dimers to sub-microkelvin temperatures by narrow-line laser cooling. 
We can make the best use of sub-microkelvin molecules by building a molecular fountain setup\cite{demi08}, where narrowing the spectral linewidth to about 1 Hz is expected. 
Another solution to achieve a narrower linewidth is holding the molecules in an optical lattice with a magic wavelength\cite{Ye08}.

By means of these techniques, spectral linewidths of about 1 Hz will be achieved. In addition, by making the efforts for a larger number of molecules and more data accumulations, the improvement of the precision by two-three orders of magnitude is expected in future experiments.
Though the uncertainty of the calculation of BBR shift gives the biggest systematic error in our results, the BBR shift coefficients can be experimentally determined with good accuracy by the measurements in the long-wavelength infrared fields like CO$_2$ lasers, as performed for the ion clock transition\cite{Bar18}.

\section*{Methods}

\subsection*{Ultracold molecular spectroscopy}
Our experimental procedure for producing KRb molecules is described in detail in \cite{aika10}(Fig.\ref{fig:schematic}). 
Briefly, KRb molecules in the $|X^1\Sigma^+, v=91, N=0\rangle$ state were produced by photoassociation in a magneto-optical trap (MOT) of $^{87}$Rb and $^{41}$K atoms. 
The temperature of the molecules was about 140 ${\mu}$K, which is close to the temperature of atoms.
Then the MOT field is turned off and both the Rb and K atoms were blasted off by shining the resonant lasers in order to suppress the energy shift caused by the molecule-atom collisions.
After that, molecules were transferred to the initial state ({\it i.e.}, $v = 86$ state in the $X^1\mathrm{\Sigma^+}$ potential) by stimulated Raman adiabatic passage (STIRAP). We used an intermediate state for the Raman transition of $|(3)^1\Sigma^+, v=41, J=1\rangle$ with Raman laser wavelengths of 875 nm and 872 nm. 
After irradiating the molecules with a microwave pulse, molecules in the initial state were optically pumped to different levels by applying the 872 nm laser.
Then the molecules in the final state were selectively observed by resonance enhanced multiphoton ionization (REMPI) by applying a pulsed laser whose wavenumber was 16218.1 cm$^{-1}$.

\subsection*{Labeling of the hyperfine states}
We express the hyperfine states based on Hund's case $b_{\beta S}$ \cite{bbs}.
The hyperfine states were labeled $|S, I_\mathrm{Rb}, F_1, I_\mathrm{K}, F_2, N, F, m_F\rangle$, where
$S$ is the total electron spin, $I_{\mathrm{Rb}}$($I_{\mathrm K}$) is the nuclear spin of rubidium (potassium),
$N$ is the rotational angular momentum, and $F$($m_F$) is (the $z$ component of) the total angular momentum of the molecule.
$F_1$ and $F_2$ are defined as
$\mathbf{F_\mathrm{1}\equiv S+I_\mathrm{Rb}}$ and $\mathbf{F_\mathrm{2}\equiv F_\mathrm{1}+I_\mathrm{K}}$, and $F$ satisfies the relation $\mathbf{F\equiv F_\mathrm{2}+N}$.
Note that $I_\mathrm{Rb}=I_\mathrm{K}=3/2$. 
We only discuss $N=0$ states in this letter; thus, $F=F_2$.
In the Results section, we express the hyperfine states as $|S, F_1, F, m_F\rangle$ in short.

\subsection*{Uncertainty of the sensitivity}
We calculated the sensitivities of $|\rm{i}\rangle$ and $|\rm{f}\rangle$ by using the experimentally determined potential energy curves reported in \cite{pash07}.
Their uncertainties were estimated from the uncertainties of the vibrational level intervals \cite{demi08}, which were estimated from the difference between the vibrational energies predicted by \cite{pash07} and our experimental observations.

\section*{Data availability}
The data that support the plots within this paper and other findings of this study are available from the corresponding author upon reasonable request.

\bibliography{apssamp}

\section*{End Notes}
\subsection*{Acknowledgements}
This work was supported by a Grant-in-Aid for Young Scientists (A) of JSPS (No. 23684034), a Grant-in-Aid for Scientific Research on Innovative Areas of JSPS (No. 24104702), JST PRESTO Grant Number JPMJPR17P5, Japan and the Matsuo Foundation.

\subsection*{Author contributions}
J. K. and A. O. performed the experiment and analyzed the
data. 
J. K. and S.I. supervised the project and prepared the manuscript. 

\subsection*{Competing financial interests}
The authors declare no competing interests.

\end{document}


\title{Supplementary Information for ``Measurement of the variation of electron-to-proton mass ratio using
ultracold molecules produced from laser-cooled atoms''}

\author{Kobayashi et al.}

\maketitle

\newpage

\noindent
{{\bf Supplementary note 1}}

As given in Supplementary table 1, we observed four singlet and six triplet states by STIRAP spectroscopy.
The hyperfine structures were analyzed based on the Hamiltonian 
\begin{equation}
H = A_{\mathrm{K}} \mathbf{S_{\mathrm{K}}\cdot I_{\mathrm{K}}}+A_{\mathrm{Rb}} \mathbf{S_{\mathrm{Rb}}\cdot I_{\mathrm{Rb}}},
\end{equation}
where $\mathbf{S_\mathrm{K}}$($\mathbf{S_\mathrm{Rb}}$) and $A_{\mathrm{K}}$($A_{\mathrm{Rb}}$) are the electron spin and hyperfine coupling constant for the K(Rb) atom, respectively. 
The fitting parameters are the energy separation ($E_0$) between $|X^1\Sigma^+, v=86, N=0\rangle$ and $|a^3\Sigma^{+}, v=16, N=0\rangle$ states without the hyperfine interaction and hyperfine coupling constants ($A_\mathrm{K}$ and $A_\mathrm{Rb}$) for both atoms.
In the analysis, we ignore mixing with higher rotational states because it is less important.
The fitting agrees well with the experimental data as shown in Supplementary table 1. 
From the analysis, we can predict the singlet-triplet mixing, the Zeeman coefficients, and the sensitivity to the variation of $\mu$ as shown in Supplementary table 1.

\newpage

\renewcommand{\thetable}{{\bf{\arabic{table}}}}
\renewcommand{\tablename}{{\bf Supplementary table}}
\begin{table}[h]
\begin{center}

\begin{tabular}{c c c c c c c c}\hline\hline
	&$F_1$	& $F$		& $E_{\mathrm{exp}}$	& $E_{\mathrm{cal}}$&$\rho_X$ 	& $\epsilon_{\rm{B}}$	& $W$ \\
	&		& 			& (MHz)					& (MHz)				&(\%)		&(kHz\,$\mu T^{-2}$)	& (GHz)		\\ \hline
\multirow{10}{1.1cm}{$a^3\Sigma^+$,\\$v = 16$,\\$S = 1$}
	& \multirow{4}{0.5cm}{5/2}
			& 4			& - 					& 7448.16			& 0.0000	& 0.693			& 1783	\\ \cline{3-8}
	& 		& 3			& - 					& 7349.17			& 0.0009	& 0.880			& 1783	\\ \cline{3-8}
	& 		& 2			& - 					& 7273.68			& 0.0008	& 1.128			& 1783	\\ \cline{3-8}
	&	 	& 1			& - 					& 7222.73			& 0.0004	& -2.475		& 1783	\\ \cline{2-8}
	& \multirow{4}{0.5cm}{3/2}
			& 3			& 3231.81 				& 3231.87			& 1.93		& 0.558			& 1970	\\ \cline{3-8}
	& 		& 2			& 3190.73 				& 3190.76			& 2.09		& 0.961			& 1985	\\ \cline{3-8}
	& 		& 1			& 3160.29 				& 3160.28			& 2.22		& 3.239			& 1997	\\ \cline{3-8}
	& 		& 0			& 3144.01 				& 3143.94			& 2.29		& -4.649		& 2004	\\ \cline{2-8}
	& \multirow{2}{0.5cm}{{\bf 1/2}}
			& {\bf 1}			&{\bf 634.94} 				& {\bf 634.87}			& {\bf 0.113}		&{\bf  0.806}			& {\bf 1794}	\\ \cline{3-8}
	& 		& 2			& 550.20 				& 550.26			& 0.254		& -1.108		& 1807	\\ \hline
\multirow{4}{1.1cm}{$X^1\Sigma^+$,\\$v = 86$,\\$S = 0$}
	& \multirow{4}{0.5cm}{{\bf 3/2}}
			& 3			& 12.29 				& 12.30				& 98.07		& -0.0084		& 11283	\\ \cline{3-8}
	& 		& 2			& 5.10 					& 5.14				& 97.65		& -0.0076		& 11242	\\ \cline{3-8}
	& 		& 1			& 1.54 					& 1.56				& 97.67		& -0.0022		& 11243	\\ \cline{3-8}
	& 		& {\bf 0}			& {\bf 0} 			& {\bf 0}		&{\bf 97.71}		& {\bf -0.0164}		& {\bf 11248}	\\ \hline\hline
\end{tabular}
\caption{\label{tbl:1} Hyperfine states of the $|X^1\Sigma^+, v = 86\rangle$ and $|a^3\Sigma^+, v = 16\rangle$ states. 
The hyperfine states are denoted as $|S, F_1, F, m_F\rangle$.
$E_{\mathrm{exp}}$ and $E_{\mathrm{cal}}$ are energies obtained from the experiment and the data fitting, respectively. Both are measured from the $|S = 0, F_1=3/2, F=0\rangle$ state. 
Singlet components ($\rho_X$), second order Zeeman shift coefficients for $m_F=0$ states($\epsilon_{\rm{B}}$), and the sensitivity to the variation of the electron-to-proton mass ratio ($W\equiv \partial E/\partial(\mathrm{ln}\,\mu)$) are also shown.
The obtained value for $E_0$ (4720.27 MHz) is consistent with the predicted value (4178.75 MHz) in Ref.\cite{pash07}.
The obtained values for $A_\mathrm{K}$ (125.54 MHz) and $A_\mathrm{Rb}$ (3384.99 MHz) are slightly smaller than the hyperfine coupling constants for bare atoms({\it i.e.},127.007 MHz and 3417.341 MHz, respectively \cite{arim77}) and result from the dependence of the coupling constants on the internuclear distance.
The transition between $|0, 3/2, 0, 0\rangle$ and $|1, 1/2, 1, 0\rangle$ states, indicated by boldface type, was used to test the stability of $\mu$.
}
\end{center}
\end{table}

\newpage

\bibliography{apssamp}